\documentstyle[a4p,11pt,oldlfont,epsfig]{article}

\def\Gcs{{{\mathrm GeV}/c}^{2}}
\def\Gc{{{\mathrm GeV}/c}}
\def\Mcs{{{\mathrm MeV}/c}^{2}}

\def\dedx{{\mathrm d}E/{\mathrm d}x}
\def\Z{{\mathrm Z}^{0}}
\def\Bc{\mathrm B_c}

\begin{document}
\begin{titlepage}
\begin{center}{\large   EUROPEAN LABORATORY FOR PARTICLE PHYSICS
}\end{center}\bigskip
\begin{flushright}
       CERN-PPE/97-137\\ \today
\end{flushright}
\bigskip\bigskip\bigskip\bigskip\bigskip
\begin{center}{\LARGE\bf Search for the $\mathbf{B_c}$ Meson in
    Hadronic ${\mathbf Z^0}$ Decays}
\end{center}\bigskip\bigskip
\def\thefootnote{\ast}
 \begin{center}{\LARGE The OPAL Collaboration
}\end{center}\bigskip\bigskip
\bigskip\begin{center}{\large  Abstract}\end{center}

\noindent A search for decays of the $\Bc$ meson was performed using data
collected from 1990--1995 with the OPAL detector on or near the $\Z$ peak at
LEP. The decay channels $\Bc^+\rightarrow
{\mathrm J}/\psi\pi^+$, 
$\Bc^+\rightarrow {\mathrm J}/\psi {\mathrm a}_1^+$ and  $\Bc^+\rightarrow
{\mathrm J}/\psi\ell^+\nu$ were investigated, where $\ell$ 
denotes an electron or a muon.
Two candidates are observed in the mode 
$\Bc^+\rightarrow {\mathrm J}/\psi\pi^+$, with an
estimated background of $(0.63\pm 0.20)$ events. The weighted mean of
the masses
of the two candidates is $(6.32\pm 0.06)$~$\Gcs$, which is consistent with
the predicted mass of the $\Bc$ meson.
One candidate event is observed in the mode 
$\Bc^+\rightarrow {\mathrm J}/\psi\ell^+\nu$, with an estimated
background of $(0.82\pm 0.19)$ events.
No candidate events are observed in the
$\Bc^+\rightarrow {\mathrm J}/\psi {\mathrm a}_1^+$ decay mode, with
an estimated background of $(1.10\pm 0.22)$ events.
Upper
bounds at the 90\% confidence level are set on the production rates for these
processes.\\

\bigskip\bigskip\bigskip
\bigskip\bigskip\bigskip\bigskip
\bigskip\bigskip
\begin{center}{\large
(Submitted to Physics Letters B)
}\end{center}
\end{titlepage}

\begin{center}{\Large        The OPAL Collaboration
}\end{center}\bigskip
\begin{center}{
K.\thinspace Ackerstaff$^{  8}$,
G.\thinspace Alexander$^{ 23}$,
J.\thinspace Allison$^{ 16}$,
N.\thinspace Altekamp$^{  5}$,
K.J.\thinspace Anderson$^{  9}$,
S.\thinspace Anderson$^{ 12}$,
S.\thinspace Arcelli$^{  2}$,
S.\thinspace Asai$^{ 24}$,
S.F.\thinspace Ashby$^{  1}$,
D.\thinspace Axen$^{ 29}$
% ...
}\end{center}\bigskip
$^{  1}$School of Physics and Space Research, University of Birmingham,
Birmingham B15 2TT, UK
\newline
$^{  2}$Dipartimento di Fisica dell' Universit\`a di Bologna and INFN,
I-40126 Bologna, Italy
\newline
% ...
$^{  5}$Cavendish Laboratory, Cambridge CB3 0HE, UK
\newline
% ...
$^{  8}$CERN, European Organisation for Particle Physics,
CH-1211 Geneva 23, Switzerland
\newline
$^{  9}$Enrico Fermi Institute and Department of Physics,
University of Chicago, Chicago IL 60637, USA
\newline
% ...
$^{ 12}$Indiana University, Department of Physics,
Swain Hall West 117, Bloomington IN 47405, USA
\newline
% ...
$^{ 16}$Department of Physics, Schuster Laboratory, The University,
Manchester M13 9PL, UK
\newline
% ...
$^{ 23}$Department of Physics and Astronomy, Tel Aviv University,
Tel Aviv 69978, Israel
\newline
$^{ 24}$International Centre for Elementary Particle Physics and
Department of Physics, University of Tokyo, Tokyo 113, and
Kobe University, Kobe 657, Japan
\newline
% ...
$^{ 29}$University of British Columbia, Department of Physics,
Vancouver BC V6T 1Z1, Canada
\newline
% ...
$^{  a}$ and at TRIUMF, Vancouver, Canada V6T 2A3
\newline
% ...
$^{  c}$ and Institute of Nuclear Research, Debrecen, Hungary
\newline
% ...
\newpage

\section{Introduction}

\begin{figure}[htbp]
 \begin{center}
   \mbox{\epsfxsize=16cm\epsffile{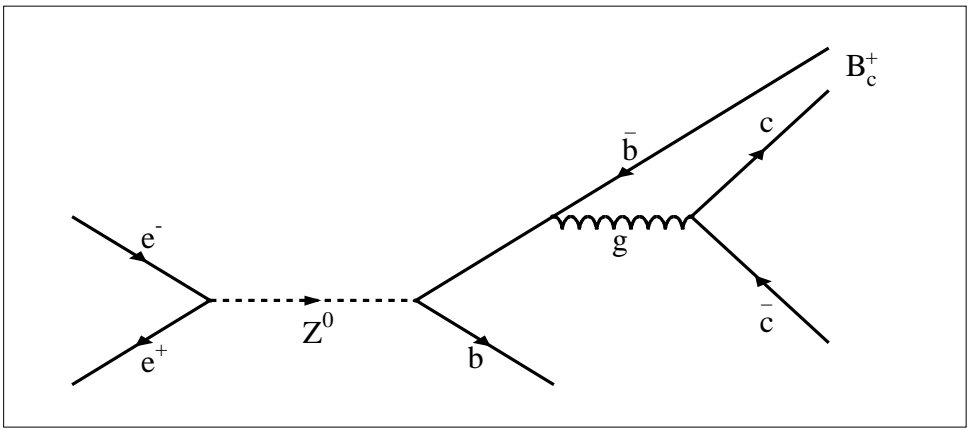}}
  \end{center}
  \caption{Feynman Diagram for the predicted production process 
$\Z \rightarrow {\mathrm b\bar{b}} \rightarrow \Bc X$ }   
  \label{fig:bc-prod-diagram}
\end{figure}

The ground-state pseudoscalar mesons containing a $\mathrm b$ quark 
have all been observed, by experiments at CESR,
DORIS, LEP and the TEVATRON, except for the beauty-charm meson 
$\Bc {\mathrm (\bar{b}c)}$.  The $\Bc$ meson can be produced at LEP in
hadronic $\Z$ decays.
Using the non-relativistic potential model for heavy quark bound states, the 
mass of the $\Bc$
meson is predicted to be in the range 6.24 to 6.31 
$\Gcs$~\cite{bc_mass_eichten}. The 
production mechanism for the ${\mathrm \bar{b}c}$ bound
states differs from that of the 
$\mathrm B_d$, $\mathrm B_u$ and $\mathrm B_s$ mesons, since the soft
fragmentation process, involving spontaneous creation of 
${\mathrm b\bar{b}}$ or
${\mathrm c\bar{c}}$, is severely suppressed. The predicted 
dominant production mechanism shown in
figure~\ref{fig:bc-prod-diagram} involves the emission and splitting
to ${\mathrm c\bar{c}}$ of a hard gluon in the process  
$\Z\rightarrow{\mathrm b\bar{b}}$~\cite{bc_prod_chang}. Perturbative 
QCD calculations predict a production rate
of $10^{-5}$ to $10^{-4}$ $\Bc$ per hadronic $\Z$ decay, with a  momentum
spectrum that is considerably softer than that of the lighter $\mathrm B$ 
hadrons~\cite{bc_prod_braaten} (see
figure~\ref{fig:mom_spectrum}). The 
decay of
the $\Bc$ meson is governed by the weak interaction; strong decay into a 
lower mass beauty meson and a
charmed hadron is forbidden by energy conservation.
There is a large
spread in the predictions for the $\Bc$ lifetime, although it is generally 
agreed that it
is shorter than the lifetime of the light $\mathrm B$ mesons. 
Theoretical calculations predict a 
significant branching ratio into modes
involving the ${\mathrm J}/\psi$ meson~\cite{theo_psibr}. 

In this article we report on a search for $\Bc$ decays in a data sample of
$4.2\times10^{6}$ hadronic $\Z$ decays collected with the OPAL detector at
LEP. A previous article by the OPAL collaboration~\cite{psi_paper}
reported on a study of ${\mathrm J}/\psi$ meson production in 
hadronic $\Z$ decays,
and the reconstruction of exclusive decays of $\mathrm B$ 
hadrons into modes containing a
${\mathrm J}/\psi$ meson. This analysis included a candidate for the 
decay~\footnote{Throughout this article charge conjugate modes are implied.}
$\Bc^+\rightarrow {\mathrm J}/\psi\pi^+$. The analysis of the $\Bc$ 
decays is extended to include searches for the decay
modes $\Bc^+\rightarrow {\mathrm J}/\psi {\mathrm a}_1^+$ and
$\Bc^+\rightarrow {\mathrm J}/\psi \ell^+\nu$, as well as
$\Bc^+\rightarrow {\mathrm J}/\psi \pi^+$, where $\ell$ 
denotes an electron or a muon.\\
                          
\section{The OPAL Detector}
 
The OPAL detector is described in detail elsewhere~\cite{opald}. Here we
briefly describe the components which are relevant to this analysis. 
The OPAL coordinate
system is defined with the $z$-axis following the electron beam
direction, the $x$-axis pointing towards the center of the LEP ring and
the $y$-axis pointing upwards, forming a right-handed coordinate system.
The polar angle $\theta$ is defined relative to the $z$-axis, and $r$
and $\phi$ are the standard cylindrical polar coordinates.
Charged particle tracking is performed by the 
central detector system that is located
in a solenoidal magnetic field of 0.435~T. The central tracking system consists
of a two layer silicon microvertex detector, installed before the 1991
run~\cite{opal_si}, a high precision
vertex drift chamber, a large volume jet chamber and a set of planar drift
$z$-chambers
measuring track coordinates along $z$. The 
momentum resolution
of the central detector in the $x$-$y$ plane is
$(\delta p_{xy}/p_{xy})^2=(2\%)^2+(0.15\%\cdot p_{xy})^2$, 
where $p_{xy}$ is  in
$\Gc$. Particle identification is provided by the measurement of
specific ionization, $\dedx$, in the jet chamber. The $\dedx$ resolution for
tracks with the maximum of 159 samplings 
in the jet chamber is 3.5\%~\cite{dedx}. The central
detector is surrounded by a lead glass electromagnetic calorimeter with a
pre-sampler. The magnet yoke is instrumented with layers of streamer tubes
that serve as a hadron calorimeter and provide additional information for
muon identification. Four layers of planar drift chambers surrounding the
detector provide tracking for muons.

\section{Data Sample and Event Selection}
 
The data used in this study were collected between 1990 and 1995 
using the OPAL detector at LEP.  The sample corresponds to
approximately $4.2\times 10^6$ hadronic $\Z$ decays. The
selection of hadronic events has been described elsewhere~\cite{opal_hadronic}.
The selection 
efficiency for the multihadronic events is  $(98.1 \pm 0.5) \%$, 
with a background of less than 0.1\%. 
 
For this analysis we impose the following additional requirements on charged
tracks: the number of hits in the central detector used for the reconstruction
of a track must be greater than 40 (this restricts the acceptance to
$|\cos\theta|<0.94$); the distance 
of closest approach to the beam axis in the
$x$-$y$ plane must be less than 0.5 cm; the transverse 
momentum with respect to the beam direction
must exceed 0.25 $\Gc$; and the total 
momentum of the track must exceed 0.5 $\Gc$.
To obtain accurate polar angle measurements, a barrel
($|\cos\theta|<0.72$) track is
required to match with a $z$-chamber track segment containing at least 3 hits;
forward going tracks are constrained to the point where they leave
the chamber.  

A track is identified as a pion if
the ${\mathrm d}E/{\mathrm d}x$ 
probability for the pion hypothesis, that is the probability that 
the specific ionization energy loss in the 
jet chamber $({\mathrm d}E/{\mathrm d}x)$ is compatible with that
expected for a pion, exceeds 2.5\%
if the measured ${\mathrm d}E/{\mathrm d}x$ is lower than the expected 
${\mathrm d}E/{\mathrm d}x$ for a pion, and 0.1\% if it is higher. 
For the purpose of background rejection,
tracks are identified as kaons if the $\dedx$ probability for the kaon
hypothesis is greater than 5\%.
 
Leptons are identified by imposing the following selection criteria.
We require the track momentum $p~>~2.0~\Gc$, 
and $|\cos\theta|<0.9$.  For electron
identification we use a neural network 
algorithm~\cite{nn5} which uses twelve variables containing information 
from the central tracking system, the electromagnetic 
calorimeter and its pre-sampler. The overall efficiency for
the identification of electrons 
from ${\mathrm B}$ hadron decays is 
$(77~\pm~5)\%$.  The error in the electron identification efficiency
was determined by comparing the efficiencies in Monte Carlo and data
for a pure sample of electrons from photon conversions.
For muon identification, two sets of
selection criteria are used.  For muon candidates combined to
form ${\mathrm J}/\psi$ candidates in the
$\Bc^+ \rightarrow {\mathrm J}/\psi \pi^+$ and 
$\Bc^+ \rightarrow {\mathrm J}/\psi a_1^+$ modes
we employ a ``normal'' muon selection.
In this selection we require a $\phi$-$\theta$ match between the 
extrapolated muon candidate track and
a track segment reconstructed in the muon chamber~\cite{muon_id1}. In
addition, we require that the candidate muon track be the best match to the
muon segment. When no match to a muon segment is found, we
search for a match with a track segment in the hadron 
calorimeter~\cite{muon_idhcal}. 
The efficiency for this ``normal'' muon identification is $(85~\pm~4)\%$. 
The errors for the muon identification were determined by comparing
the efficiency between Monte Carlo and data for a pure sample of muons
from muon pair events.
For any muon candidates 
combined to form $\Bc$ candidates in the decay 
$\Bc^+ \rightarrow {\mathrm J}/\psi \ell^+\nu$, where 
${\mathrm J}/\psi \rightarrow \ell^+\ell^-$, we employ a
``strong'' muon identification~\cite{muon_id2}.  In this selection 
we use only tracks matched 
with track segments in the muon detector, reject tracks 
identified as kaons using $\dedx$ information, and apply an isolation 
cut by requiring that there
be less than 20 track segments in the muon detector
within 0.3 radians of the track. 
The efficiency for this ``strong'' muon identification used in the 
$\Bc \rightarrow {\mathrm J}/\psi \ell\nu$ mode is $(76~\pm~4)\%$.
 
Events are organised into jets of 
particles which are constructed using charged 
tracks and neutral clusters that are not associated to any charged 
track~\cite{jetfinding}. To
form jets we use the scaled invariant mass jet-finding algorithm of JADE
with a jet resolution parameter $y_{\mathrm cut}=0.04$.        

\begin{figure}[p]
  \begin{center}
   \mbox{\epsfxsize=16cm\epsffile{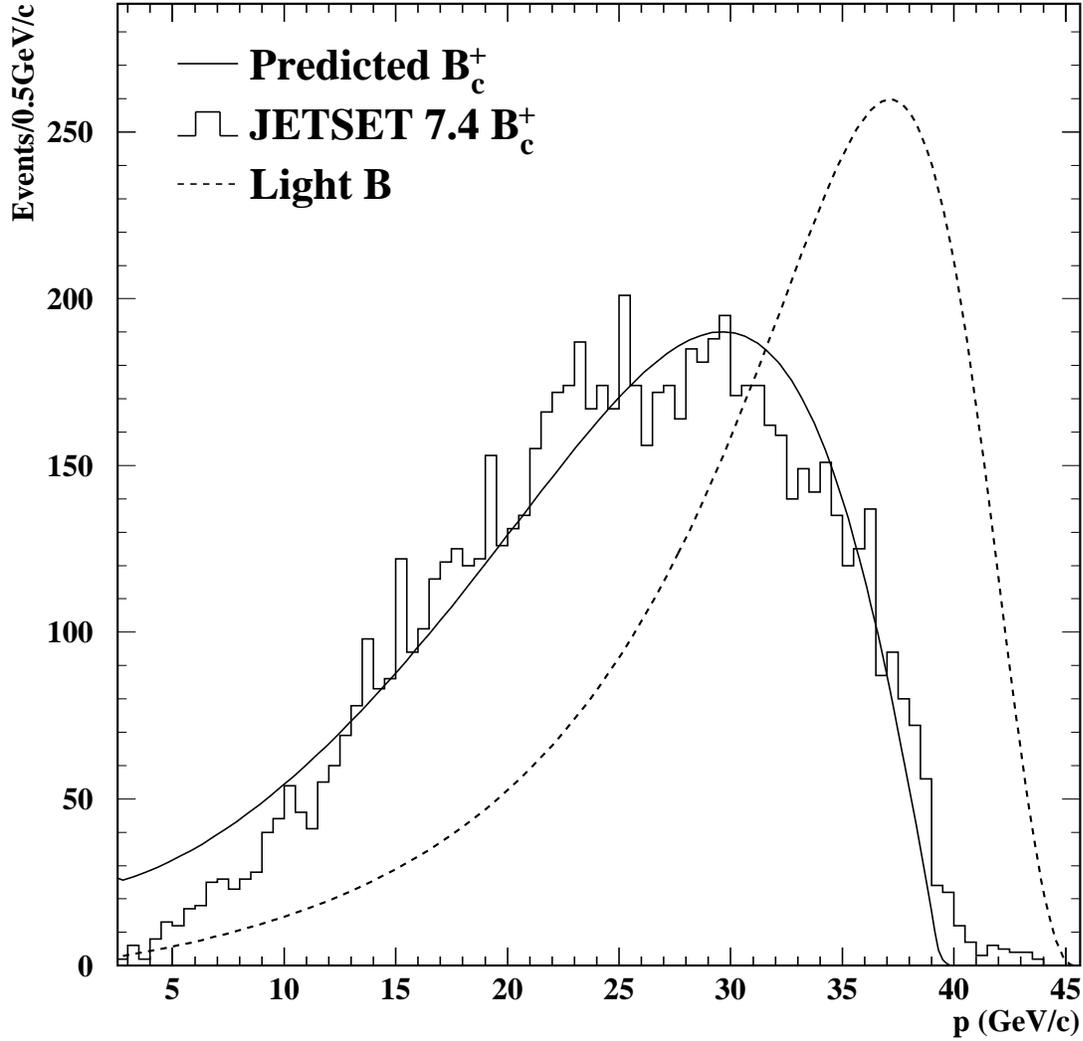}}
  \end{center}
  \caption{Momentum spectrum of $\Bc$ mesons from the process
$\Z \rightarrow {\mathrm b\bar{b}} \rightarrow \Bc X$ generated
using JETSET 7.4 (histogram).  
Overlaid is the prediction of the theoretical model in reference~[2]
(solid line).
Also shown is the momentum distribution of light $\mathrm B$ mesons
for the OPAL tune of JETSET 7.4 in which the mean energy of ${\mathrm b}$
hadrons agrees with the experimental value (dashed line).}
  \label{fig:mom_spectrum}
\end{figure}

A Monte Carlo simulation is used to determine the reconstruction and selection
efficiencies for the various decay modes and for estimating the background
level. 
The process $\Z \rightarrow {\mathrm b\bar{b}} \rightarrow \Bc X$ 
(figure~\ref{fig:bc-prod-diagram}) and subsequent $\Bc$ meson decays are
simulated using the JETSET 7.4 program~\cite{jetset}. 
Figure~\ref{fig:mom_spectrum} shows the prediction of
reference~\cite{bc_prod_chang} for the $\Bc$ momentum spectrum along with the 
JETSET 7.4 simulation of the $\Bc$ spectrum. In addition to simulating $\Bc$
production as described in reference~\cite{bc_prod_chang}, JETSET 7.4 
includes contributions from the production of
excited states of ${\mathrm \bar{b}c}$ bound states. 
The two momentum distributions are similar. Also shown is the distribution 
for light $\mathrm b$ hadrons given by the
Peterson et al. fragmentation function~\cite{peterson}
with its parameter tuned to produce the 
measured mean energy fraction 
($\langle x_E \rangle=E_{\mathrm B}/E_{\mathrm beam}$), 
indicating that the $\Bc$ spectrum is
predicted to be considerably softer than that for the light 
$\mathrm b$ hadrons. 
The measured mean
energy fraction ($\langle x_E \rangle$) 
for the light $\mathrm b$ hadrons is 
$\langle x_E \rangle= 0.695~\pm~0.006~\pm~0.008$~\cite{b_frag},
while for the generated $\Bc$ meson we find  $\langle x_E \rangle=0.54$.  
Samples of 2000 events
were simulated for each of the following decay modes: 
$\Bc^+ \rightarrow {\mathrm J}/\psi \pi^+$, $\Bc^+\rightarrow
{\mathrm J}/\psi {\mathrm a}_1^+$, 
where ${\mathrm a}_1^+\rightarrow\rho^0\pi^+$ and $\rho^0\rightarrow
\pi^+\pi^-$, and the semileptonic mode
$\Bc^+\rightarrow {\mathrm J}/\psi \ell^+\nu$, where $\ell$ denotes
an electron or a muon.  In each event the 
${\mathrm J}/\psi$ decays to $\ell^+\ell^-$.

A simulated event sample of $4\times 10^6$ five-flavor
hadronic $\Z$ decays, nearly
equal in size to the data sample, was used for studying the background
processes.  A sample of 80,000 
hadronic $\Z$ decays containing the process 
${\mathrm B} \rightarrow {\mathrm J}/\psi {\mathrm X}$, 
where a ${\mathrm J}/\psi\rightarrow \ell^+\ell^-$
decay is present in each event, was used to increase the statistical 
significance of the background study.  In addition two samples of 4000
events containing the processes
$\Z \rightarrow {\mathrm J}/\psi {\mathrm q\bar{q}}$ and 
$\Z \rightarrow {\mathrm J}/\psi {\mathrm c\bar{c}}$
were produced in order to study background due to prompt 
${\mathrm J}/\psi$ production from gluon fragmentation and 
${\mathrm c}$~quark fragmentation, respectively.
The JETSET 7.4 parton shower Monte Carlo
generator is used for the simulation of
the hadronic $\Z$ decays. For
the fragmentation of heavy quarks into charmed and light b-flavored hadrons, 
we use the Peterson fragmentation function. 
JETSET 7.4 parameters and branching ratios were tuned
to match experimental results~\cite{opal_jetset_tune}~\cite{PDG_1994}. 
All simulated events are passed through the full 
simulation of the OPAL detector~\cite{opalmc}.  JETSET does not
include radiative decay of ${\mathrm J}/\psi$ into lepton pairs.  The
presence of unreconstructed final state radiation~(FSR) in the decay
${\mathrm J}/\psi \rightarrow \ell^+\ell^-\gamma$ produces a tail
toward lower masses in the invariant mass distribution.
The effect of FSR on the ${\mathrm J}/\psi$
mass distribution is included in Monte Carlo events at
reconstruction level.  The photon energy is
calculated using first order perturbative QED~\cite{fsr_first}.
An error is calculated to account for the higher order 
terms~\cite{fsr_second}. Excepting the FSR correction for Monte Carlo
simulated events, data and Monte Carlo simulated samples are analysed 
using the same reconstruction program. 

\section{Search for $\Bc$ decays}

We search for the decay modes 
$\Bc^+ \rightarrow {\mathrm J}/\psi \pi^+$, $\Bc^+\rightarrow
{\mathrm J}/\psi {\mathrm a}_1^+$, 
where ${\mathrm a}_1^+\rightarrow\rho^0\pi^+$ and $\rho^0\rightarrow
\pi^+\pi^-$, and the semileptonic mode
$\Bc^+\rightarrow {\mathrm J}/\psi \ell^+\nu$.  
The analysis involves the reconstruction of
${\mathrm J}/\psi$ candidates in the leptonic mode 
${\mathrm J}/\psi\rightarrow\ell^+\ell^-$, which
are then combined with other tracks to form  
${\mathrm J}/\psi\pi^+$,
${\mathrm J}/\psi\pi^+\pi^-\pi^+$, 
and ${\mathrm J}/\psi\ell^+$ combinations. The ${\mathrm J}/\psi\ell^+$
combination is a partial reconstruction of the mode 
$\Bc^+\rightarrow {\mathrm J}/\psi\ell^+\nu$.  Candidates are formed
from charged tracks which are assigned to the same jet.

\subsection{Reconstruction of ${\mathrm J}/\psi$ Decays}

\begin{figure}[p]
  \begin{center}
   \mbox{\epsfxsize=16cm\epsffile{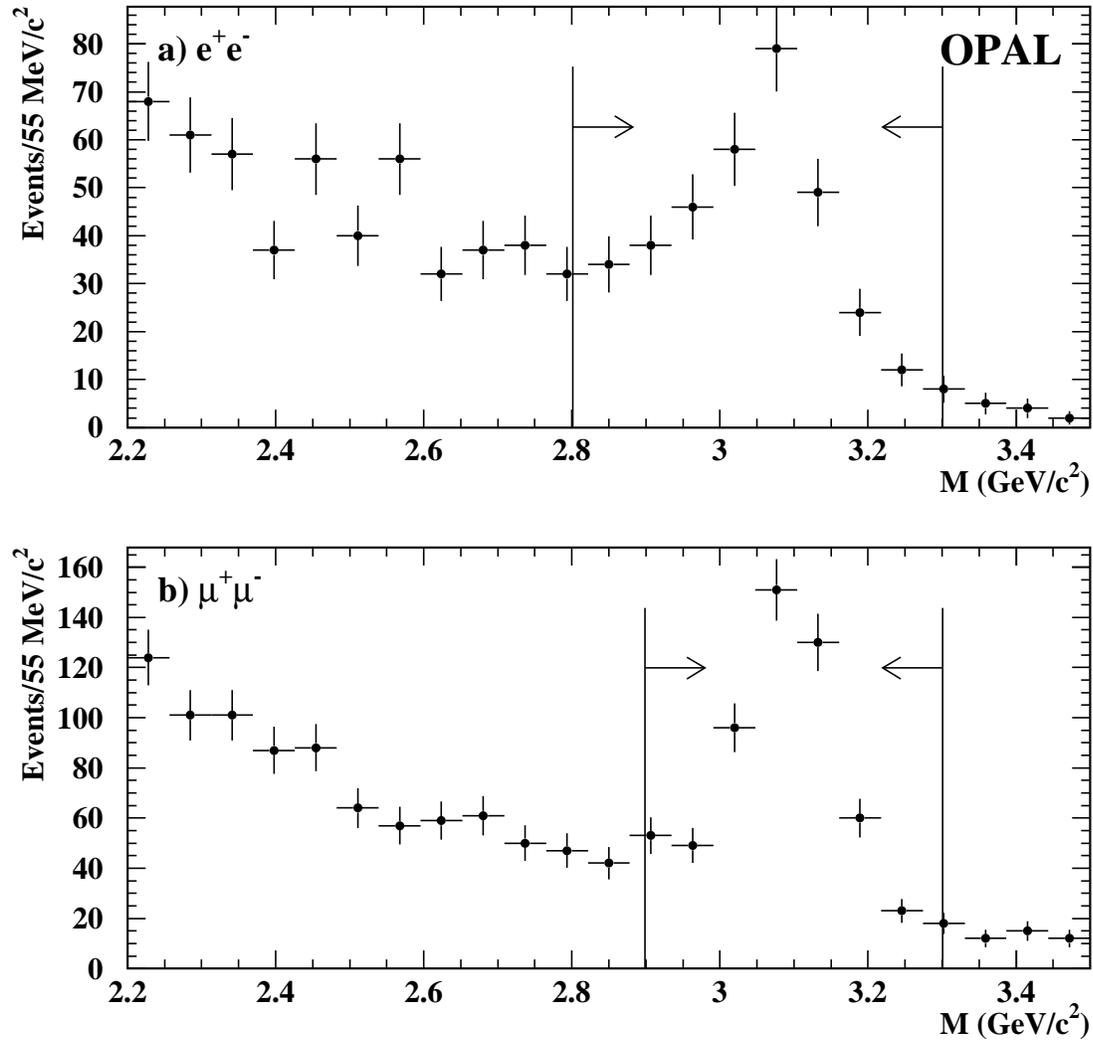}}
  \end{center}
  \caption{Invariant mass spectrum of selected 
(a) ${\mathrm e}^+ {\mathrm e}^-$, and (b) $\mu^+\mu^-$
pairs.  Also shown are the mass regions where the ${\mathrm J}/\psi$
candidates are defined.}
  \label{fig:psi}
\end{figure}

The ${\mathrm J}/\psi$ meson decays are reconstructed in the leptonic modes
${\mathrm J}/\psi\rightarrow\mu^+\mu^-$ 
and  ${\mathrm J}/\psi\rightarrow {\mathrm e}^+ {\mathrm e}^-$. A pair of 
opposite sign electron or muon candidates in the same jet, with an invariant
mass consistent with the ${\mathrm J}/\psi$ 
mass, is considered  a ${\mathrm J}/\psi$ candidate.  
Muon candidates combined to form ${\mathrm J}/\psi$ candidates in the
$\Bc^+\rightarrow {\mathrm J}/\psi \pi^+$ and
$\Bc^+\rightarrow {\mathrm J}/\psi {\mathrm a}_1^+$ modes
must satisfy the ``normal'' muon identification.
Muon candidates combined to form ${\mathrm J}/\psi$ candidates in the
$\Bc^+\rightarrow {\mathrm J}/\psi \ell^+\nu$ mode must satisfy the 
criteria of the ``strong'' muon identification.
${\mathrm J}/\psi$ candidates are required to have an  invariant
mass within the range 2.9 to 3.3~$\Gcs$ for 
the $\mu^+\mu^-$ channel, and within 
the range 2.8 to 3.3~$\Gcs$ for the ${\mathrm e}^+ {\mathrm e}^-$ 
channel. 
The ${\mathrm J}/\psi \rightarrow {\mathrm e}^+ {\mathrm e}^-$
candidate range is extended to lower masses in order to include the
tail in the ${\mathrm J}/\psi \rightarrow {\mathrm e}^+ {\mathrm e}^-$
invariant mass distribution due to electron bremsstrahlung radiation
in the detector and ${\mathrm J}/\psi$ radiative decays.
Figure~\ref {fig:psi} shows the lepton pair invariant mass distributions
for leptons selected with the ``normal'' lepton selection
in the range 2.5 to 3.5~$\Gcs$, 
where the ${\mathrm J}/\psi$ peak is clearly visible in both
${\mathrm e}^+ {\mathrm e}^-$ and $\mu^+\mu^-$ modes. 
The peak position and width are
consistent with the ${\mathrm J}/\psi$ 
mass and the expected resolution of the OPAL
detector. 
We find a total of 354
${\mathrm J}/\psi\rightarrow {\mathrm e}^+ {\mathrm e}^-$ 
candidates and 551 
${\mathrm J}/\psi\rightarrow\mu^+\mu^-$ candidates using the normal
muon selection.  We find 391 
${\mathrm J}/\psi\rightarrow\mu^+\mu^-$ candidates using the
``strong'' muon selection.

\subsection{Selection of $\Bc$ Candidates}

\begin{figure}[p] \begin{center}
\mbox{\epsfxsize=16cm\epsffile{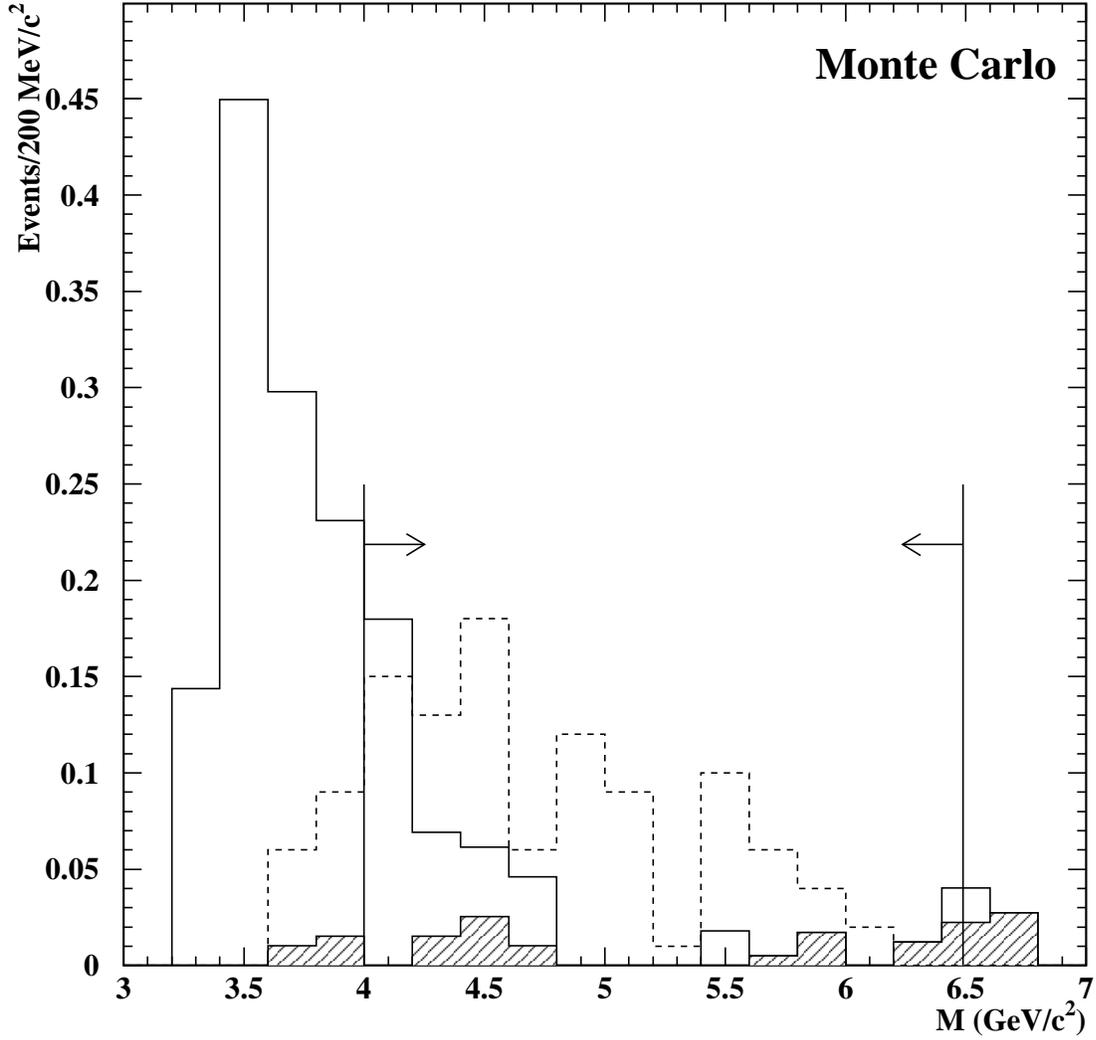}} 
\end{center}
\caption{Invariant mass distribution of reconstructed 
${\mathrm J}/\psi\ell^+$ combinations from a
simulated sample of $\Bc^+\rightarrow {\mathrm J}/\psi\ell^+\nu$ decays 
(dashed open histogram),
and combinatorial background 
from enriched samples of Monte Carlo simulated  
${\mathrm B} \rightarrow {\mathrm J}/\psi {\mathrm X}$ (solid open
histogram) 
and prompt ${\mathrm J}/\psi$ (hatched histogram) events.
The backgrounds are normalised to the number of events 
expected in the data sample.  The normalisation of the signal is
arbitrary. Also shown is the mass region where 
the $\Bc$ candidates are defined.} 
\label{fig:back-sig-semil} 
\end{figure}

The dominant background to the sample of $\Bc$ candidates is from random
combinations of ${\mathrm J}/\psi$'s produced in ${\mathrm b}$ hadron decays 
with other tracks from $\mathrm b$
hadron decays or from fragmentation.
Given the fact that at best a few events are expected in each channel, it is
crucial that the combinatorial background be reduced to below the level of the
expected signals. The selection criteria were developed by studying Monte
Carlo simulated events containing the signal processes, and the simulated
sample of five-flavor $\Z$ events (described above). In general, significant
background suppression can be achieved by taking advantage of the hard momentum
spectrum and the long lifetime of the $\mathrm b$ hadrons. 
However, the soft momentum
spectrum of the $\Bc$ weakens the discrimination power of any momentum cut.
Furthermore, since there is large uncertainty in the predictions of the $\Bc$
lifetime, no decay length cut is used. The criteria are summarised below:

\begin{itemize} 

\item[a.] In all decay modes ${\mathrm J}/\psi$ candidates
are kinematically constrained to the nominal ${\mathrm J}/\psi$ mass
in order to improve the $\Bc$ mass resolution.  

\item[b.] For the exclusive modes, 
${\mathrm J}/\psi\pi^+$ and ${\mathrm J}/\psi {\mathrm a}_1^+$,
we require that  the ${\mathrm d}E/{\mathrm d}x$ measurement for each 
pion candidate be ``consistent"
with the expected value for a pion (as described in section 3).

\item[c.] In the semileptonic mode, 
$\Bc^+ \rightarrow {\mathrm J}/\psi \ell^+\nu$, there is a large background
at lower masses
involving fake ${\mathrm J}/\psi$ or ${\mathrm J}/\psi$ combined with
fake leptons or leptons from cascade decays.
This background is reduced by using
the ``strong" muon identification
for all muon candidates and requiring the 
$\ell^+$ lepton track momentum $p>4.0$ $\Gc$.

\item[d.] For  ${\mathrm J}/\psi \pi^+\pi^-\pi^+$ 
combinations, we require that the
three-pion combination be consistent with resulting from the decay 
${\mathrm a}_1^+\rightarrow \rho^0\pi^+$, 
where $\rho^0\rightarrow\pi^+\pi^-$. The
invariant mass of the three 
pion combination must be consistent with the ${\mathrm a}_1$
mass, ($1.0<M(\pi^+\pi^-\pi^+)<1.6$)~$\Gcs$, 
and the invariant mass of at least one
of the two  $\pi^+\pi^-$ pairs must be in the $\rho^0$ mass range,
($0.65<M(\pi^+\pi^-)<0.90$)~$\Gcs$.

\item[e.] All tracks from the $\Bc$ must be consistent with originating
from the same decay vertex.
For each $\Bc$ candidate we determine the decay vertex from the
intersection of the tracks, 
including the tracks forming the ${\mathrm J}/\psi$
candidate, in the $x$-$y$ plane. We require the $\chi^2$
probability of the vertex fit to exceed 1\%.

\item[f.] Since the combinatorial 
backgrounds are largest at low momenta, we
impose a minimum momentum cut on the $\Bc$ 
candidates. For the ${\mathrm J}/\psi\ell^+\nu$ mode, 
where the full momentum of the candidate is not reconstructed,
we require the momentum of the ${\mathrm J}/\psi\ell^+$ combination
to exceed 30\% of the beam energy.
For ${\mathrm J}/\psi\pi^+$ combinations 
we require the momentum of the $\Bc$ candidate
to exceed 55\% of the beam energy. For the 
$\Bc^+ \rightarrow {\mathrm J}/\psi {\mathrm a}_1^+$ candidates,
where the combinatorial background is more severe, the candidate momentum is
required to exceed 70\% of the beam energy.

\item[g.] For the exclusive modes,  we 
take advantage of the fact that the decay
products of a pseudoscalar meson
are isotropically distributed in its rest frame. 
We define $\theta^*$ as the angle between the $\Bc$
candidate direction and the direction of the 
${\mathrm J}/\psi$ in the $\Bc$ rest frame.  Since the
combinatorial background mainly peaks near the backward direction 
($\cos\theta^*=-1.0$), we
require $\cos\theta^*>-0.8$.

\item[h.] For the exclusive modes, the invariant mass of the $\Bc$ candidates
must be in the mass interval 6.0 
to 6.5~$\Gcs$ (hereafter referred to as the
signal region) which is centered around the predicted $\Bc$ mass, with a width
which is about three times the $\Bc$ 
mass resolution ($\simeq~80~\Mcs$) on each side.  The mass window includes
the entire range of predictions for the $\Bc$ mass.
For the semileptonic mode, where 
the full invariant mass cannot be calculated, the invariant mass of the
${\mathrm J}/\psi\ell^+$ combination is 
used to define the signal region. The rarity
of random combinations of three leptons in hadronic $\Z$ decays
combined with the high mass of the $\Bc$ produces a natural separation point
between the signal and the background combinations. 
Figure~\ref{fig:back-sig-semil} shows the invariant mass distribution for
${\mathrm J}/\psi\ell^+$ combinations from 
a sample of simulated  $\Bc^+\rightarrow
{\mathrm J}/\psi\ell^+\nu$ decays, along 
with the distribution of the combinatorial
background from simulated samples of the 
${\mathrm B} \rightarrow {\mathrm J}/\psi {\mathrm X}$ and prompt 
${\mathrm J}/\psi$ events (see the following section for detail).  The
signal distribution peaks above 4.0 $\Gcs$ and the background combinations are 
mostly at lower masses. Hence we restrict 
the signal region to the mass interval 4.0 to 6.5 $\Gcs$.

\item[i.]${\mathrm B}^+ \rightarrow {\mathrm J}/\psi {\mathrm K}^+$ decays can
fake $\Bc^+\rightarrow {\mathrm J}/\psi\ell^+\nu$ decays if the kaon
is misidentified as a lepton.
For the semileptonic mode we reject candidates that have a reconstructed 
mass within $3 \sigma$ in mass resolution ($\sigma~\simeq~60~\Mcs$)
of the measured ${\mathrm B^+}$ mass when the
third lepton candidate is assigned the kaon mass.

\end{itemize}

\subsection{Estimation of Reconstruction Efficiencies and the 
Background Levels}

The reconstruction efficiencies are calculated from Monte Carlo generated event
samples, with the $\Bc$ meson simulated at a mass of $6.25~\Gcs$.
In the ${\mathrm J}/\psi\pi^+$ mode the leptons from the ${\mathrm J}/\psi$
decay are expected to have a $\sin{^2 \theta}$ angular distribution
with respect to the ${\mathrm J}/\psi$ direction in the $\Bc$
rest-frame.  In order to simulate this distribution, which was not
included in the Monte Carlo generator, the selected events in the 
${\mathrm J}/\psi\pi^+$ mode were reweighted.  
For the decay $\Bc^+\rightarrow {\mathrm J}/\psi {\mathrm a}_1^+$ we
conservatively assume an ${\mathrm a}_1^+$ width of 400~MeV.
Each sample is composed of events with the appropriate $\Bc$ decay and
subsequent ${\mathrm J}/\psi \rightarrow \ell^+\ell^-$ decay. 
The mass resolution
is found to be about 80 $\Mcs$ in the modes 
${\mathrm J}/\psi\pi^+$ and ${\mathrm J}/\psi {\mathrm a}_1^+$.
The reconstruction efficiencies for these modes are $(10.0~\pm~0.7)\%$ and
$(1.8~\pm~0.3)\%$, respectively, and $(5.5~\pm~0.5)\%$ for 
the semileptonic mode
$\Bc^+\rightarrow {\mathrm J}/\psi \ell^+ \nu$, where the errors are 
due to Monte Carlo statistics.  The reconstruction 
efficiency is sensitive to the $\Bc$
momentum distribution.  An estimate of this sensitivity is found
by comparing the efficiencies obtained 
using the distribution predicted by JETSET 7.4 with 
those obtained assuming the theoretical 
calculations of reference~\cite{bc_prod_chang}.  Values of 
${\mathrm m_b} = (4.9~\pm~0.2)~\Gcs$ and 
${\mathrm m_c} = (1.5~\pm~0.2)~\Gcs$ were used for the input quark 
masses~\cite{bc_prod_braaten}. 
In the 
${\mathrm J}/\psi\pi^+$ mode the difference is 16.3\%.  For the
${\mathrm J}/\psi {\mathrm a}_1^+$ mode, 
where a harder cut of $x_E > 0.7$ is applied,
a difference of 37.9\% is found.  In the semileptonic mode the
difference is
5.7\%.  We take the systematic errors on the
efficiencies to be one half of the difference for each mode.

The expected background level in each channel
is determined by searching for $\Bc$
decays in the simulated hadronic $\Z$ event sample. 
According to Monte Carlo simulations the background combinations at masses 
below the signal region are dominated by the
combinations of a real ${\mathrm J}/\psi$ with random tracks from
$\mathrm b$ hadron decays or from
the fragmentation processes.  
The ${\mathrm J}/\psi$ mesons originate dominantly from the
decays of $\mathrm b$ 
hadrons, with a  small fraction, 
$(4.8~\pm~1.7~\pm~1.7)\%$~\cite{prompt_psi_paper}, 
of prompt ${\mathrm J}/\psi$ (${\mathrm J}/\psi$ resulting 
from the fragmentation processes). In total, the 
simulated hadronic event sample contains 2800 events
containing a leptonic ${\mathrm J}/\psi$ 
decay. To improve the statistical
significance of the background studies we apply the $\Bc$ search to a sample of
80,000 hadronic $\Z$ decays containing the process 
${\mathrm B} \rightarrow {\mathrm J}/\psi {\mathrm X}$, 
where a ${\mathrm J}/\psi\rightarrow \ell^+\ell^-$
decay is present in each event. This Monte Carlo sample is equivalent to 30
times the size of the data sample. 
In addition, the $\Bc$ search was applied to two samples of 4000
events containing the processes
$\Z \rightarrow {\mathrm J}/\psi {\mathrm q\bar{q}}$ and 
$\Z \rightarrow {\mathrm J}/\psi {\mathrm c\bar{c}}$
in order to study background due to prompt 
${\mathrm J}/\psi$ production from gluon fragmentation and 
${\mathrm c}$~quark fragmentation, respectively, which are 
the predicted dominant production mechanisms for 
prompt ${\mathrm J}/\psi$~\cite{prompt_psi_theo}.  
The predicted branching ratios for these processes are 
${\mathrm Br}(\Z \rightarrow {\mathrm J}/\psi {\mathrm q\bar{q}})~=~
1.9~\times~10^{-4}$ and
${\mathrm Br}(\Z \rightarrow {\mathrm J}/\psi {\mathrm c\bar{c}})~=~
0.8~\times~10^{-4}.$
These Monte Carlo sets are equivalent to 40 and 100 times the size of the
data sample, respectively.

\begin{figure}[p]
  \begin{center}
   \mbox{\epsfxsize=16cm\epsffile{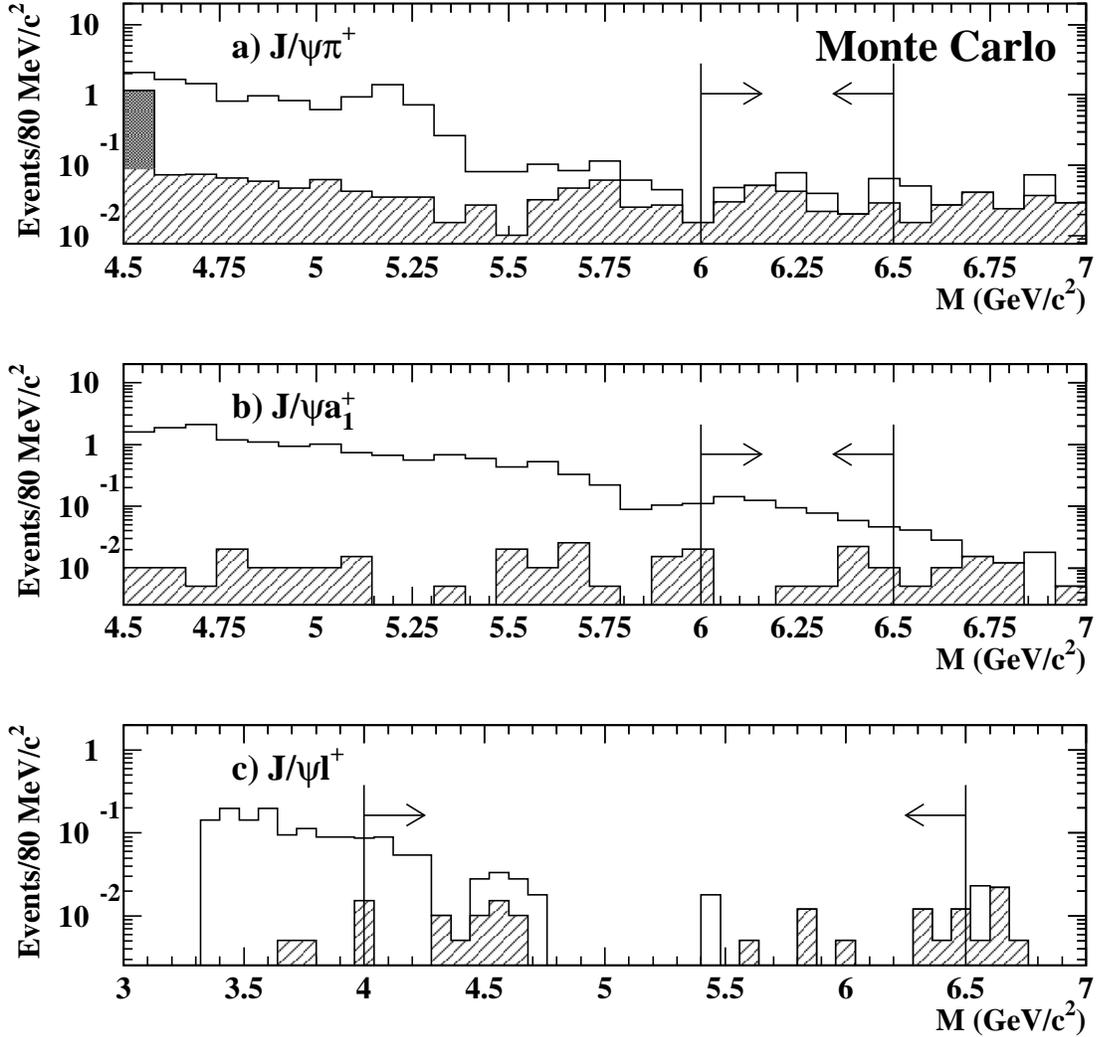}}
  \end{center}
  \caption{Invariant mass distribution 
of (a) ${\mathrm J}/\psi \pi^+$ combinations,
(b) ${\mathrm J}/\psi {\mathrm a}_1^+$ combinations 
and (c) ${\mathrm J}/\psi\ell^+$ combinations in
the simulated background samples.
Background events from the process 
${\mathrm B} \rightarrow {\mathrm J}/\psi {\mathrm X}$
are represented by the open histogram.  Background events from prompt  
${\mathrm J}/\psi$ processes are represented by the hatched histogram.
Background events involving fake ${\mathrm J}/\psi$ are represented by
the shaded histogram.
Also shown are the mass regions where the $\Bc$ candidates are defined.}
  \label{fig:mc-back-mass2}
\end{figure}

The resulting invariant mass distributions 
for the $\Bc$ candidates in the three modes are shown in 
figure~\ref{fig:mc-back-mass2}.  From these 
distributions we estimate the background
level (see table~\ref{table:back}) by counting the number of 
candidates in the signal region and normalising to the number of events 
expected in the data sample.  The contribution to the background from
${\mathrm B} \rightarrow {\mathrm J}/\psi {\mathrm X}$  
events is normalised using the measured rate 
${\mathrm Br}(\Z \rightarrow {\mathrm J}/\psi {\mathrm X})~=~
(3.9~\pm~0.2~\pm~0.3)~\times~10^{-3}$~\cite{psi_paper}.  
The contribution from prompt ${\mathrm J}/\psi$ is normalised using
the measured rate of prompt ${\mathrm J}/\psi$ production
${\mathrm Br}(\Z \rightarrow~{\mathrm prompt}~{\mathrm J}/\psi {\mathrm X})~=~
(1.9~\pm~0.7~\pm~0.7)~\times~10^{-4}$, with the two components of the
prompt ${\mathrm J}/\psi$ signal each assigned a fraction of the total
branching rate according to the 
ratio of their theoretical production rates.  The measured rate of
prompt ${\mathrm J}/\psi$ production is in agreement with the
theoretical rates.
The uncertainty in the normalisation factors 
is used as a systematic error for each background estimate.
 
\begin{table}[h]
\begin{center}
\begin{tabular}{|c||c|c|c|} \hline
Background & \multicolumn{2}{c}{~~~~~~~~~~~~~~~~~~~~~~~~~~~~~Decay Mode}&\\ \hline
& $\Bc^+\rightarrow {\mathrm J}/\psi\pi^+$ 
& $\Bc^+\rightarrow {\mathrm J}/\psi {\mathrm a}_1^+$
& $\Bc^+\rightarrow {\mathrm J}/\psi\ell^+\nu$\\ \hline
${\mathrm B} \rightarrow {\mathrm J}/\psi {\mathrm X}$ 
& $0.22~\pm~0.09~\pm~0.02$ &  $1.01~\pm~0.19~\pm~0.09$ 
& $0.61~\pm~0.15~\pm~0.06$\\ 
${\mathrm J}/\psi$(c quark frag.) 
& $0.24~\pm~0.05~\pm~0.13$ &  $0.07~\pm~0.03~\pm~0.04$ 
& $0.13~\pm~0.04~\pm~0.07$\\ 
${\mathrm J}/\psi$(gluon frag.) 
& $0.17~\pm~0.06~\pm~0.09$ &  $0.02~\pm~0.02~\pm~0.01$ 
& $0.07~\pm~0.04~\pm~0.04$\\ \hline
Total 
& $0.63~\pm~0.12~\pm~0.16$ &  $1.10~\pm~0.19~\pm~0.10$ 
& $0.82~\pm~0.16~\pm~0.10$\\ \hline
\end{tabular}
\end{center}
\caption{Background estimates from samples of 80,000 
${\mathrm B} \rightarrow{\mathrm J}/\psi {\mathrm X}$  
enriched events, 4000 prompt ${\mathrm J}/\psi$ from the gluon
fragmentation process, $\Z \rightarrow {\mathrm J}/\psi {\mathrm q\bar{q}}$,
and 4000 prompt ${\mathrm J}/\psi$ from the ${\mathrm c}$~quark
fragmentation process, 
$\Z \rightarrow {\mathrm J}/\psi {\mathrm c\bar{c}}$.
The first error is statistical and the second is
systematic.}
\label{table:back}
\end{table}

\section{Results and Upper Bounds on the Production Rates}

\begin{figure}[p]
  \begin{center}
   \mbox{\epsfxsize=16cm\epsffile{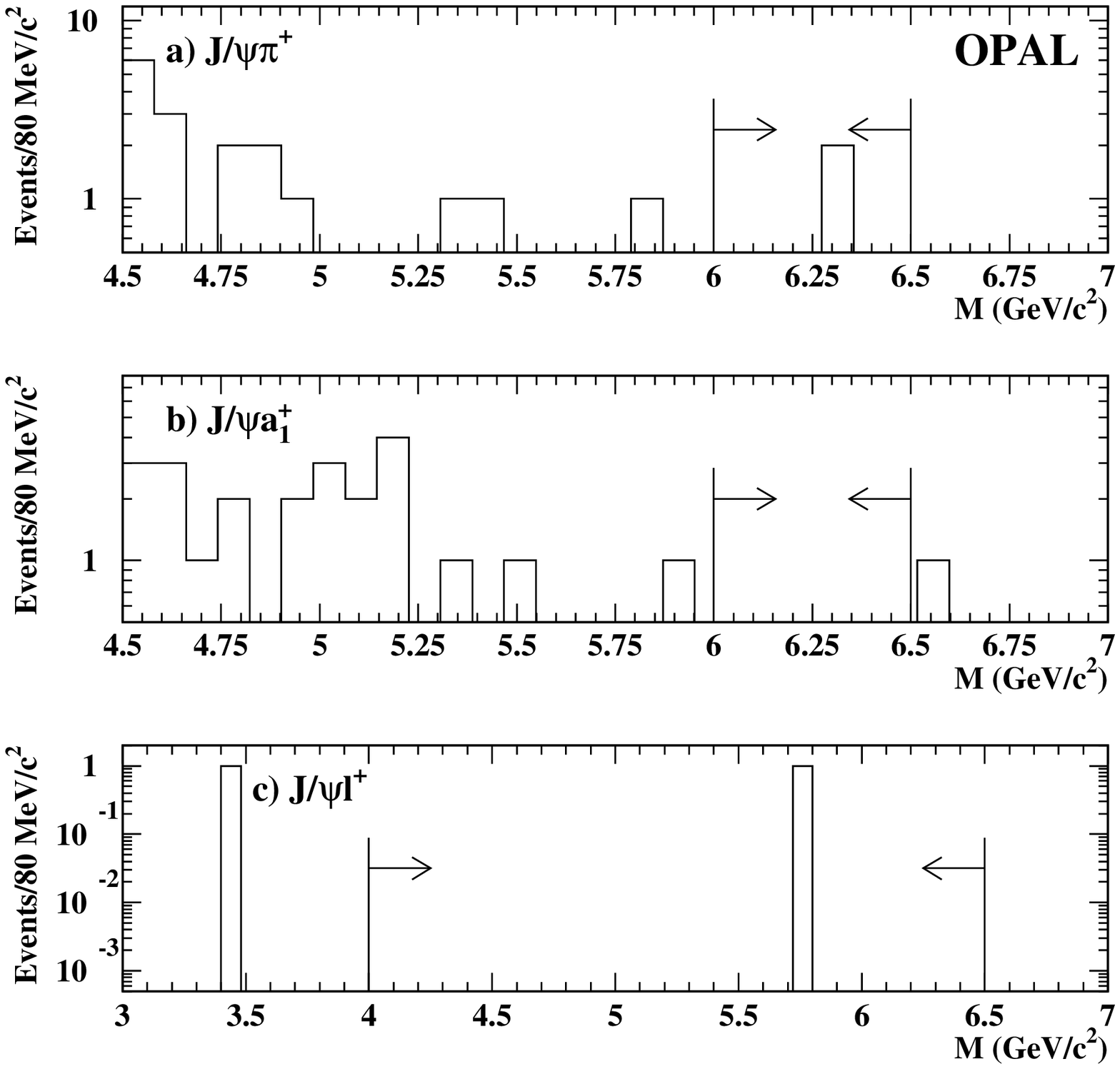}}
  \end{center}
  \caption{Invariant mass distribution 
of (a) ${\mathrm J}/\psi\pi^+$ combinations,
(b) ${\mathrm J}/\psi {\mathrm a}_1^+$ combinations
and (c) ${\mathrm J}/\psi\ell^+$ combinations in the data sample.
Also shown are the mass regions where the $\Bc$ candidates are defined.}
  \label{fig:bc-candidates}
\end{figure}

Figure~\ref{fig:bc-candidates} shows the invariant mass distributions of
the ${\mathrm J}/\psi\pi^+$, 
${\mathrm J}/\psi {\mathrm a}_1^+$, 
and ${\mathrm J}/\psi \ell^+\nu$ candidates in
the data sample. The
shapes and overall levels of the distributions below the signal regions are
consistent with the distributions obtained from the simulated background
samples. In the mode ${\mathrm J}/\psi\pi^+$ 
we find two events in the signal range 6.0 to
6.5 $\Gcs$. The invariant masses of the 
two $\Bc$ candidates  are $(6.29~\pm~0.17)$~$\Gcs$ 
and $(6.33~\pm~0.063)$~$\Gcs$.  The errors are calculated from the
errors on the track parameters.
The reconstructed decay times of the
two candidates are $\tau = (-0.06~\pm~0.19)$~ps and 
$\tau = (0.09~\pm~0.10)$~ps,
respectively. The estimated background in this mode is
$(0.63~\pm~0.20)$ events. 
The probability for a background of $0.63$ events to fluctuate to 2 or
more is 13.2\%.
In the mode ${\mathrm J}/\psi \ell^+\nu$ we find one event in the signal
region.  The mass of the candidate ${\mathrm J}/\psi \ell^+$
combination is $5.76$~$\Gcs$.
The momentum is $15.4$~$\Gc$.
The reconstructed decay length is $(0.14~\pm~0.14)$~cm.  The estimated
background in this mode is $(0.82~\pm~0.19)$ events.  
In the signal region
above the mass of the candidate event, $5.76$ to $6.5~\Gcs$, the 
probability to observe one or more events from background is 8.9\%, while the
efficiency is reduced to $(0.6\pm 0.2)\%$.
There are no
candidate events in the ${\mathrm J}/\psi {\mathrm a}_1^+$ mode in the
signal region compared with an estimated background of
$(1.10~\pm~0.22)$ events.

We determine an upper bound at 90\% confidence 
level on the number of events in
each channel by applying Poisson 
statistics to the number of events observed in
the signal region, without background subtraction.
This is used to calculate an  upper limit on the production rate from:

\begin{eqnarray}
\label{rate1}
{{\mathrm Br}(\Z \rightarrow \Bc^+ {\mathrm X}) \over 
{\mathrm Br}(\Z \rightarrow {\mathrm q\bar{q}})}
~\times~
{\mathrm Br}(\Bc^+\rightarrow final~state)~=
~{\mathit N}(at~90\%~C.L.)~
\times~{\epsilon_{had} \over ({\mathit N}(\Z)\times \epsilon\times
{\mathrm Br}({\mathrm J}/\psi\rightarrow \ell^+\ell^-))}, \nonumber
\end{eqnarray}\\

\noindent
where $N(\Z)$ is the total number of hadronic $\Z$ events in the data sample;
$\epsilon_{had}$ is the hadronic event selection efficiency, $0.981~\pm~0.005$;
$\epsilon$ is the reconstruction efficiency for each mode; and
${\mathrm Br}({\mathrm J}/\psi\rightarrow \ell^+\ell^-)$ is
the branching ratio for the leptonic decays of 
${\mathrm J}/\psi$, ${\mathrm J}/\psi\rightarrow  {\mathrm e}^+ {\mathrm e}^-$
and ${\mathrm J}/\psi\rightarrow \mu^+\mu^-$, $0.1203~\pm~0.0028$~\cite{pdg}. 
For the  mode $\Bc^+\rightarrow {\mathrm J}/\psi {\mathrm a}_1^+$, 
we also account for the branching ratio
${\mathrm Br}({\mathrm a}_1^+\rightarrow\rho^0\pi^+)~=~0.5$.
The total systematic uncertainties on the branching ratio upper limits
for each mode are shown in table~\ref{table:branch_sys_errors}.
These systematic uncertainties are included in the following
90\% confidence level upper limits using the technique 
of ref.~\cite{sys_error}:

\begin{table}[t!]
\begin{center}
\begin{tabular}{|c||c|c|c|} \hline
Error Source & \multicolumn{2}{c}{~~~~~~~~~~~~~~~~~~~~~~~~Decay Mode}&\\ \hline
& $\Bc^+\rightarrow {\mathrm J}/\psi\pi^+$ 
& $\Bc^+\rightarrow {\mathrm J}/\psi {\mathrm a}_1^+$
& $\Bc^+\rightarrow {\mathrm J}/\psi\ell^+\nu$\\ \hline
MC statistics         & 7.2\% & 16.6\% & 9.1\% \\ 
$\Bc$ fragmentation   & 8.2\% & 19.0\% & 2.9\% \\ 
Lepton eff.           & 6.5\% & 6.5\% & 7.5\% \\ 
${\mathrm Br}({\mathrm J}/\psi)$ & 2.3\% & 2.3\% & 2.3\%\\
hadronic eff.         & 0.5\% & 0.5\% & 0.5\%\\
final state radiation & 0.6\% & 0.6\% & 0.6\% \\ \hline
Total error           & 12.8\% & 26.2\% & 12.4\% \\ \hline
\end{tabular}
\end{center}
\caption{Summary of systematic errors on the product branching 
ratio upper limits.}
\label{table:branch_sys_errors}
\end{table}

\begin{eqnarray}
{{\mathrm Br}(\Z \rightarrow \Bc^+ {\mathrm X}) \over 
{\mathrm Br}(\Z \rightarrow {\mathrm q\bar{q}})}
~\times~
{\mathrm Br}(\Bc^+\rightarrow {\mathrm J}/\psi\pi^+)~<~1.06\times 10^{-4},
\nonumber
\end{eqnarray}

\begin{eqnarray}
{{\mathrm Br}(\Z \rightarrow \Bc^+ {\mathrm X}) \over 
{\mathrm Br}(\Z \rightarrow {\mathrm q\bar{q}})}
~\times~
{\mathrm Br}(\Bc^+\rightarrow {\mathrm J}/\psi 
{\mathrm a}_1^+)~<~5.29\times 10^{-4},
\nonumber
\end{eqnarray}

\begin{eqnarray}
{{\mathrm Br}(\Z \rightarrow \Bc^+ {\mathrm X}) \over 
{\mathrm Br}(\Z \rightarrow {\mathrm q\bar{q}})}
~\times~
{\mathrm Br}(\Bc^+\rightarrow {\mathrm J}/\psi\ell^+\nu)
~<~6.96\times 10^{-5},
\nonumber
\end{eqnarray}\\

\noindent
where the branching ratio for the mode 
$\Bc^+\rightarrow {\mathrm J}/\psi\ell^+\nu$ is for decay to either
${\mathrm J}/\psi\ {\mathrm e}^+\nu$ or ${\mathrm J}/\psi\mu^+\nu$.\\

If we interpret the two candidate events in the 
$\Bc^+\rightarrow {\mathrm J}/\psi\pi^+$ mode as signal we find a
branching ratio of,\\

\begin{eqnarray}
{{\mathrm Br}(\Z \rightarrow \Bc^+ {\mathrm X}) \over 
{\mathrm Br}(\Z \rightarrow {\mathrm q\bar{q}})}
~\times~
{\mathrm Br}(\Bc^+\rightarrow {\mathrm
  J}/\psi\pi^+)~=~(3.8^{+5.0}_{-2.4}~\pm~0.5)\times 10^{-5},
\nonumber
\end{eqnarray}

\noindent
where the first error is statistical and the second systematic.

\section{Conclusion}

We have performed a search for 
$\Bc$ meson decays in data collected with the OPAL detector at 
LEP.  Two candidate $\Bc \rightarrow {\mathrm J}/\psi\pi^+$ decays 
are observed in a mass window around the
theoretical prediction of the $\Bc$ mass, compared with an estimated
background of $(0.63\pm 0.20)$ events in this mode.  The weighted average
mass of the two candidates is $(6.32\pm 0.06)$~$\Gcs$, which is
consistent with the predicted mass of the $\Bc$ meson.
One candidate is observed in the mode 
$\Bc^+\rightarrow {\mathrm J}/\psi\ell^+\nu$,
with a ${\mathrm J}/\psi\ell^+$ mass of $5.76~\Gcs$.
The estimated background in this mode is $(0.82\pm 0.19)$ events.
We have also searched for 
the decay $\Bc^+\rightarrow {\mathrm J}/\psi {\mathrm a}_1^+$,
but no candidate events were observed.  
The estimated background in this mode is $(1.10\pm 0.22)$ events.
Upper limits
at the 90\% confidence level are calculated for the production rates
of these processes,

\begin{eqnarray}
{{\mathrm Br}(\Z \rightarrow \Bc^+ {\mathrm X}) \over 
{\mathrm Br}(\Z \rightarrow {\mathrm q\bar{q}})}
~\times~
{\mathrm Br}(\Bc^+\rightarrow {\mathrm J}/\psi\pi^+)~<~1.06\times 10^{-4},
\nonumber
\end{eqnarray}

\begin{eqnarray}
{{\mathrm Br}(\Z \rightarrow \Bc^+ {\mathrm X}) \over 
{\mathrm Br}(\Z \rightarrow {\mathrm q\bar{q}})}
~\times~
{\mathrm Br}(\Bc^+\rightarrow {\mathrm J}/\psi 
{\mathrm a}_1^+)~<~5.29\times 10^{-4},
\nonumber
\end{eqnarray}

\begin{eqnarray}
{{\mathrm Br}(\Z \rightarrow \Bc^+ {\mathrm X}) \over 
{\mathrm Br}(\Z \rightarrow {\mathrm q\bar{q}})}
~\times~
{\mathrm Br}(\Bc^+\rightarrow {\mathrm J}/\psi\ell^+\nu)
~<~6.96\times 10^{-5}.
\nonumber
\end{eqnarray}

The branching ratio limits for the $\Bc^+\rightarrow {\mathrm J}/\psi\pi^+$
and $\Bc^+\rightarrow {\mathrm J}/\psi\ell^+\nu$ mode are comparable with
the limits reported by the ALEPH, DELPHI and CDF 
collaborations~\cite{aleph}\cite{delphi}\cite{CDF}.

\section{Acknowledgements}

We particularly wish to thank the SL Division for the efficient operation
of the LEP accelerator at all energies
 and for
their continuing close cooperation with
our experimental group.  We thank our colleagues from CEA, DAPNIA/SPP,
CE-Saclay for their efforts over the years on the time-of-flight and trigger
systems which we continue to use.  In addition to the support staff at our own
institutions we are pleased to acknowledge the  \\
Department of Energy, USA, \\
National Science Foundation, USA, \\
Particle Physics and Astronomy Research Council, UK, \\
Natural Sciences and Engineering Research Council, Canada, \\
Israel Science Foundation, administered by the Israel
Academy of Science and Humanities, \\
Minerva Gesellschaft, \\
Benoziyo Center for High Energy Physics,\\
Japanese Ministry of Education, Science and Culture (the
Monbusho) and a grant under the Monbusho International
Science Research Program,\\
German Israeli Bi-national Science Foundation (GIF), \\
Bundesministerium f\"ur Bildung, Wissenschaft,
Forschung und Technologie, Germany, \\
National Research Council of Canada, \\
Research Corporation, USA,\\
Hungarian Foundation for Scientific Research, OTKA T-016660, 
T023793 and OTKA F-023259.\\


\begin{thebibliography}{3}
 
\bibitem{bc_mass_eichten}
E. J. Eichten and C. Quigg, Phys. Rev.  {\bf D49} (1994) 5845;\\
S. S. Gershtein, V. V. Kiselev, A. K. Likhoded, and A. V. Tkabladze, Phys. Rev.
{\bf D51} (1995) 3613.

\bibitem{bc_prod_chang}
Chao-Hsi Chang and Yu-Qi Chen, Phys. Rev. {\bf D46} (1992) 3845.

\bibitem{bc_prod_braaten}
E. Braaten and K. Cheung, Phys. Rev. {\bf D48} (1993) 5049.

\bibitem{theo_psibr}
S. S. Gershtein, A. K. Likhoded, and S. R. Slabospitsky, Int. J. Mod. Phys.
{\bf 6} (1991) 2309.

\bibitem{psi_paper}
OPAL Collab., G. Alexander et al., Z. Phys. {\bf C70} (1996) 197.

\bibitem{opald}
OPAL Collaboration, K. Ahmet et al., Nucl. Instrum. and Meth. {\bf A305}
(1991) 275.
 
\bibitem{opal_si}
P. P. Allport et al., Nucl. Instrum. and Meth. {\bf A324} (1993) 34;\\
P. P. Allport et al., Nucl. Instrum. and Meth. {\bf A346} (1994) 476.

\bibitem{dedx}
M. Hauschild et al., Nucl. Instrum. and Meth. {\bf A314} (1992) 74.

\bibitem{opal_hadronic}
OPAL Collab., G. Alexander et al., Z. Phys. {\bf C52} (1991) 175.

\bibitem{nn5}
OPAL Collab., G. Alexander et al., Z. Phys. {\bf C70} (1996) 357.

\bibitem{muon_id1}
OPAL Collab., K. Ackerstaff et al., Z. Phys. {\bf C74} (1997) 423.

\bibitem{muon_idhcal}
OPAL Collab., R. Akers et al., Z. Phys. {\bf C60} (1993) 199.

\bibitem{muon_id2}
OPAL Collab., P. Acton et al., Z. Phys. {\bf C58} (1993) 523.

\bibitem{jetfinding}
OPAL Collab., P. Acton et al., Z. Phys. {\bf C63} (1994) 197.

\bibitem{jetset}
T. Sj\"{o}strand, Comp.Phys.Comm. {\bf 39} (1986) 347;\\
M. Bengtsson and T. Sj\"{o}strand, Comp. Phys. Comm. {\bf 43} (1987) 367;\\ 
M. Bengtsson and T. Sj\"{o}strand, Nucl. Phys. {\bf B289} (1987) 810;\\
T. Sj\"{o}strand, CERN-TH/6488-92.

\bibitem{peterson}
C. Peterson, D. Schlatter, I. Schmitt and P. Zerwas, Phys. Rev. {\bf D27}
(1983) 105.

\bibitem{b_frag}
OPAL Collab., G. Alexander et al., Phys. Lett.  {\bf B364} (1995) 93.

\bibitem{opal_jetset_tune}
Parameter values of JETSET were tuned 
to describe global event shape variables:
OPAL Collab. R. Akers et al., Z. Phys. {\bf C65} (1995) 31.

\bibitem{PDG_1994}
L. Montanet et al., Phys. Rev. {\bf D50} (1994) 1.

\bibitem{opalmc}
J. Allison et al., Nucl. Instrum. and Meth. {\bf A317} (1992) 47.

\bibitem{fsr_first}
F. A. Berends and R Kleiss, Nucl. Phys. {\bf B177} (1981) 237.

\bibitem{fsr_second}
O. Nicrosini and L. Trentadue, Phys. Lett. {\bf B196} (1987) 551;\\
J. P. Alexander et al., Phys. Rev. {\bf D37} (1988) 56.

\bibitem{prompt_psi_paper}
OPAL Collab., G. Alexander et al., Phys. Lett. {\bf B384} (1996) 343.

\bibitem{prompt_psi_theo}
P. Cho, Phys. Lett. {\bf B368} (1996) 171;\\
P. Cho and A. Leibovich, Phys. Rev. {\bf D53} (1996) 150.


\bibitem{pdg}
R. M. Barnett et al., Phys. Rev. {\bf D54} (1996) 1.

\bibitem{sys_error}
R. D. Cousins and V. L. Highland, 
Nucl. Instrum. and Meth. {\bf A320} (1992) 331.

\bibitem{aleph}
ALEPH Collaboration, D. Abbaneo et al., CERN-PPE/97-026, 
submitted to Phys. Lett. {\bf B}. 

\bibitem{delphi}
DELPHI Collaboration, P. Abreu et al., CERN-PPE/96-194, 
submitted to Phys. Lett. {\bf B}. 

\bibitem{CDF}
CDF Collaboration, F. Abe et al., Phys. Rev. Lett. {\bf 77} (1996) 5176.

\end{thebibliography}
\end{document}